\documentclass[prb,aps,showpacs,preprintnumbers,amsmath,amssymb,superscriptaddress,floatfix,footinbib,twocolumn]{revtex4}
\usepackage{graphicx}
\usepackage{dcolumn}
\usepackage{bm}
\begin{document}
\title{Asymptotic self-consistency in quantum transport calculations}
\author{H. Mera}
\affiliation{Department of Physics, University of York, Heslington, York YO10
5DD, United Kingdom.}
\author{P. Bokes}
\affiliation{Department of Physics, Faculty of Electrical Engineering and Information Technology, Slovak University of Technology,
Ilkovi\v{c}ova 3, 812 19 Bratislava, Slovakia.}
\author{R.W. Godby}
\affiliation{Department of Physics, University of York, Heslington, York YO10
5DD, United Kingdom.}
\date{\today}
\begin{abstract}
\emph{Ab-initio} simulations of quantum transport commonly focus on
a central region which is considered to be connected to infinite, periodic
leads
through which the current flows.
The electronic structure of these distant leads is normally
obtained from an \emph{equilibrium} calculation, ignoring
the self-consistent response of the leads to the current.
We examine the consequences of this, and
show that the electrostatic potential, $\Delta \phi$, is
effectively being approximated by the difference between
electrochemical potentials, $\Delta \mu$, and that this approximation is
incompatible with asymptotic charge neutrality.
In a test calculation for a simple metal-vacuum-metal junction,
we find large errors in the non-equilibrium
properties calculated with this approximation,
in the limit of small vacuum gaps.
We provide a simple scheme by which these errors may be corrected.
\end{abstract}
\maketitle
\section{Introduction}
The last fifteen years have seen considerable progress in the
simulation of non-equilibrium many-electron nanoscale systems
\cite{McCannBrown,Lang1,Datta1,Taylor1,Taylor2,Palacios1,Palacios2} typically using the
Local Density Approximation (LDA) to Density Functional Theory (DFT)\cite{KS,HK}
within the framework of non-equilibrium Green's functions (NEGF) \cite{Keldysh}. 
In these simulations the systems under study consist of two electrodes (or leads),
placed to the left and to the right of an active central region which 
contains a molecule and parts of the left and right electrode. Starting from the unconnected electrode-central region-electrode 
system with each of the electrodes itself in equilibrium but not in equilibrium with each other \cite{Caroli}
the NEGF formalism then provides the formal apparatus to switch on the contacting terms 
of the Hamiltonian adiabatically, causing a current to flow through the system.
Associated with this current and 
related to the resistance of the molecule there is a ``resistivity
dipole'' arising from the newly induced charge density, that causes the electrostatic potential to drop in the neighbourhood of the central region. 
The magnitude of the drop in the self-consistent electrostatic potential is essentially fixed by a charge neutrality condition, i.e., the
fact that the asymptotic electrode regions must themselves be charge-neutral since a net charge would cause the electrostatic potential to diverge.
In the case of jellium electrodes, this charge neutrality condition acquires
a strict local form \footnote{In the case of electrodes with atomic structure the neutrality condition is quasi-local since electronic and
ionic density have to be integrated over a volume larger than $\lambda_{TF}^3$
where $\lambda_{TF}$ is the Thomas-Fermi screening length.} since, asymptotically, the electron density exactly cancels the background density at any point.

It should be pointed out that most practical 
implementations of the NEGF formalism cannot, strictly speaking, properly take into account the drop in the electrostatic potential 
between the leads, 
since the adiabatic switching of the contacting terms of the Hamiltonian (i.e. the perturbation) changes the density and the response to
that change cannot be described inside the realm of static DFT. Furthermore the lead self-energies, which describe the coupling between the
leads and the central region, are commonly obtained from an equilibrium calculation \cite{Taylor1,Taylor2}. At the level of the Hartree
approximation and in the non-equilibrium regime, this means that the leads 
do not respond to the flow of charge induced by the applied bias voltage; the electrochemical potentials remain fixed to their equilibrium
values and the drop in the self-consistent
electrostatic potential, $\Delta \phi$, is, as discussed below, effectively being approximated 
by the difference between electrochemical potentials, $\Delta \mu$ \cite{McCannBrown,Taylor1,Taylor2,Hirose}.
At the level of the Hartree-Fock approximation this lack of asymptotic self-consistency also implies that the
non-equilibrium Fock operator deep inside the leads would be equal to the equilibrium one, which is clearly not the case since the Fock
operator depends on the non-equilibrium occupancies of the current carrying states, which are different from the equilibrium ones
everywhere. 
However, we will not further discuss the effects of the lack of
         asymptotic self-consistency in the Fock operator. The most severe
         effects appear already in the self-consistent Hartree potential which
         we will use as an  illustration in our paper.
We would like to note that non-partitioned NEGF approaches,
as suggested by Cini~\cite{Cini80} and late elaborated by Stefanucci and
Almbladh~\cite{Stefanuci1,Stefanuci2}, are in theory free from these objections as they focus
on the evalualtion of the non-equilibrium Green's function in the whole
transporting system.

The relation between the electrostatic drop and asymptotic charge neutrality is already implicit in the original form of the 
Landauer formula  \cite{Landauer57,Buttiker85}
$I=\frac{T}{R}\Delta \phi$, and has been explored by some authors over the years \cite{potz,Bokes} 
until very recently \cite{Datta2}.
In Ref.~\onlinecite{Datta2} the authors further clarify the distinction between the difference between the electrochemical potentials of the left and
right electrodes, $\Delta \mu$, and the drop in the electrostatic potential, $\Delta \phi$, as well as the role played by the geometry.
However they do not discuss in detail the validity of the approximation $\Delta \mu = \Delta \phi$. 
\begin{figure*}[t]
\includegraphics[scale=0.5]{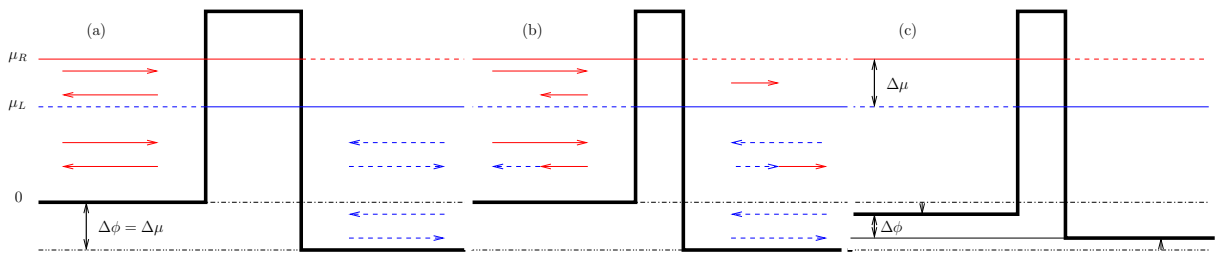}
\caption{(color online).Schematic illustration of the self-consistent electrostatic potential (thick solid line) together with its drop $\Delta \phi$, and the difference between
electrochemical potentials $\Delta \mu=\mu_R-\mu_L$. In (a) and (b) $\Delta \mu = \Delta \phi$. Solid (dashed) arrows are indicative of the
transmission and reflection amplitudes of right-going (left-going) states. In (a) a thick barrier with neglible
transmission between $\mu_L$ and $\mu_R$ is shown. In this case for the system to be neutral in its far left and right regions $\Delta \mu
=\Delta \phi$. The thinner barrier in (b) allows for larger values of the transmission coefficient and hence, since there is a net and
significant flow of particles from left to right, there local density of states is smaller on the left than on the right. If $\Delta \mu
=\Delta \phi$ this case leads to a net accumulation of charge on the right region and a depletion of it in the left region. Therefore
$\Delta \phi$ has to deviate from $\Delta \mu$ in the way schematically shown in (c).} 
\label{Fig-0}
\end{figure*}

To fix ideas consider the biased system depicted in Fig.~\ref{Fig-0}, which is translationally invariant in two of the three spatial directions and has some
localized inhomogeneity along the $z$-direction. Since left- and right-going scattering states are occupied up to two different electrochemical 
potentials, $\mu_L$ and $\mu_R$
respectively, a current flows along $z$, and therefore there is an electrostatic potential drop $\Delta \phi$. 
\emph{If we assume} that $\Delta \mu = \mu_R-\mu_L =\Delta \phi$, then it is easy to show that in the asymptotic electrode regions the electronic
density, $\rho$, is in the case of jellium electrodes, given by \cite{potz,Bokes,ferry}:
\begin{eqnarray}
\label{eq:neutral}
\rho(z\rightarrow\pm\infty)&=&\rho_B(z\rightarrow\pm\infty)\nonumber \\
			  &\pm&\int_0^{+\infty} dk_{\pm} T(E_z)[f_R(E_z)-f_L(E_z)],
\end{eqnarray}
where $\rho_B(z)$ is the background density, $k_{\pm}$ are the magnitudes of the $z$-component of the momentum 
in the asymptotic right ($+$) and left ($-$) electrode regions,
respectively; $E_z(k_{\pm})=k_{-}^2/2=k_{+}^2/2-\Delta \phi$ is the energy associated with the motion in the direction of the current;
$T(E_z)$ is the usual transmission probabilities and $f_R$ and $f_L$ are the equilibrium
Fermi-Dirac distributions for right- and left-going electrons averaged over the components of the momentum perpendicular to the direction
of the current, each of which is characterized by an electrochemical potential $\mu_{R(L)}$. The latter, as functions of $k_{-}$,
are given by \cite{potz,ferry,Bokes}:
\begin{equation}
\label{eq:fermilr}
f_{L,R}(k_{-})=\frac{1}{(2\pi)^2}(k_{L,R}^2-k_{-}^2)\Theta(k_{-}-k_{L,R}),
\end{equation}
where $k_{L,R}$ are the asymptotic Fermi wave-vectors for left- and right-going electrons respectively, in the left asymptotic region.
Therefore, when imposing $\Delta \mu = \Delta \phi$, associated with the presence of a current from, say, left to right, 
there is a charge depletion in the asymptotic left electrode
region and a charge accumulation in the asymptotic region of the right electrode \cite{potz,ferry,Bokes,Datta2}. Therefore it is clear 
that the drop
in the \emph{self-consistent} electrostatic potential is necessarily different from $\Delta \mu$. 
It is then surprising that in many
state-of-the-art \emph{ab-initio} quantum transport simulations the approximation $\Delta \mu = \Delta \phi$ is used without further
explanation  or comments on its validity\cite{McCannBrown,Taylor1,Taylor2,Hirose}. Of the few that have considered this problem let 
us mention P\"otz \cite{potz}, who introduces a drift in the electronic distribution functions of left- and right-going electrons so that the asymptotic
electrode regions remain neutral, 
Bokes and Godby \cite{Bokes} whose proposed method
is applied in this work and Di Ventra and Lang \cite{DiventraLang} who renormalize the electron densities deep 
in the jellium electrodes.

From the above given discussion it should be clear that asymptotic self-consistency (i.e. the \emph{full} non-equilibrium Green's
function of the leads) is generally needed in order to describe the drop in
the electrostatic potential correctly. The lack of asymptotic self-consistency naturally leads to the approximation $\Delta \mu = \Delta \phi$ which
is incompatible with charge neutrality in the asymptotic lead regions. The rest of this paper is organized as follows: In Section II 
we show explicitly the effect that the lack of asymptotic self-consistency has in the calculated non-equilibrium properties 
of a simple jellium metal-vacuum-metal junction. In Section III we express these ideas in the lenguage of NEGF's, showing which terms are
commonly neglected and providing a simple recipe to incorporate them for the case of jellium electrodes. We conclude in Section IV.

\section{Example: asymptotic charge neutrality in a jellium metal-vacuum-metal 
interface}
Neglecting the non-equilibrium contributions to the Green's function of the leads results,
at the level of the Hartree or DFT-LDA approximations,
in the approximation $\Delta \mu=\Delta \phi$, which in turn is incompatible with asymptotic charge neutrality in the leads. We now show
the effects of this approximation in the calculated properties of a non-equilibrium metal-vacuum-metal junction.
Even though we will use a scattering-state-based approach its equivalence with the NEGF formalism may be noted. 

The jellium model of the metal-vacuum-metal interface \cite{Nieminen, McCannBrown,OroszBalazs}
is defined in terms of the background density:
\begin{equation}
\rho_B(z)=n_0[\theta(-z)+\theta(z-L)],
\end{equation}
where $n_0=3/4 \pi r_s^3$ and $L$ is the length of the vacuum gap. For this system we solve the Kohn-Sham equations self-consistently 
using
the Perdew-Zunger \cite{PerdewZunger} parametrization of the LDA exchange-correlation potential. Historically this system was 
the first to be studied using conventional \emph{ab-initio} techniques in a non-equilibrium regime \cite{McCannBrown,OroszBalazs}, and, for
our purposes, 
constitutes a simple system for which the electrostatic effects under study arise in the most transparent manner.

In order to ensure charge neutrality in the asymptotic electrode regions, for a
given value of $\Delta \phi$ we need to find $k_R$ and $k_L$ in Eqs.~(\ref{eq:fermilr}) such that
\begin{equation}
\label{eq:asneutral}
n_0 = \rho(z \rightarrow \pm \infty),
\end{equation}
are satisfied. 
$k_R$ and $k_L$ are related to the electrode electrochemical potentials simply by
\begin{equation}
\mu_R-\mu_L=\frac{k_R^2-k_L^2}{2},
\end{equation}
Therefore, at each step of the self-consistency cycle, we solve the Poisson equation with Dirichlet boundary conditions, fixing $\Delta \phi$ and calculating
the corresponding $\Delta \mu$ that ensures that the asymptotic left and right electrode regions remain neutral.
Strictly speaking this procedure is only justified in the case (such as the metal-vacuum-metal junction) that there is a one-to-one correspondence between $\Delta
\mu$ and $\Delta \phi$, i.e., there is a one-to-one correspondence between the applied bias and the current. For this particular case our
method is equivalent to the alternative one of fixing $\Delta \mu$ and calculating $\Delta \phi$ by solving the Poisson equation with
von-Neumann boundary conditions \cite{Palacios2}.

When studying the influence of asymptotic charge neutrality in the calculated non-equilibrium properties we solve the Kohn-Sham equations
self-consistently using the procedure described by McCann and Brown \cite{McCannBrown}, with the approximation $\Delta \mu = \Delta \phi$ 
and without it, using the asymptotic neutrality condition Eq.~(\ref{eq:asneutral}). In
particular the Poisson equation is solved using Nieminen's method \cite{Nieminen,McCannBrown,Hirose}, which greatly stabilizes the iterative process and fastens the
convergence.
\begin{figure}[t]
\includegraphics[scale=0.8]{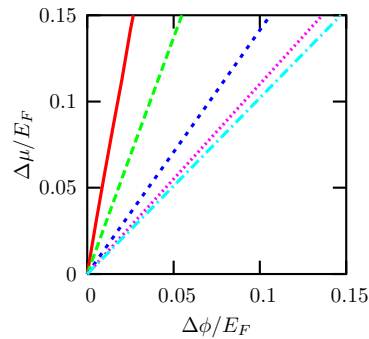}
\caption{(color online) The figure shows the difference between electrochemical potentials
$\Delta \mu$ as a
function of the drop in the electrostatic potential $\Delta \phi$, in units of
the equilibrium Fermi energy, for different values of the
electrode-electrode distance, $L$. $L=0.25 r_s$ (solid line); $L=0.5 r_s$
(dashed); $L=1r_s$
(short dashes); $L=1.5 r_s$
(dots); $L=2 r_s$
(dot-dashed). $\Delta \mu \approx \Delta \phi$ only for large
electrode-electrode spacings. For reference $E_F=3.131$ $eV$}
\label{Fig-1}
\end{figure}
\begin{figure}[t]
\includegraphics[scale=0.8]{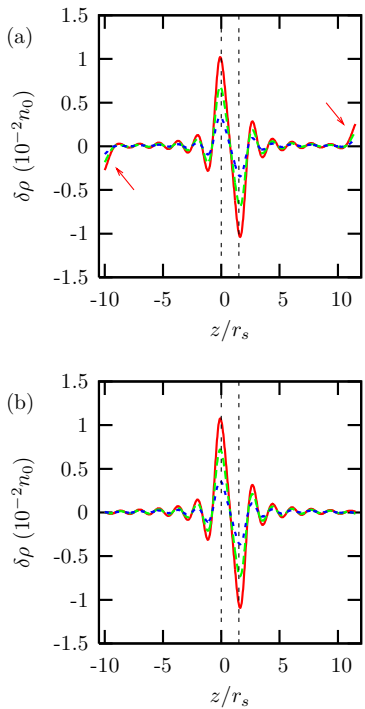}
\caption{(color online) (a) resistivity dipoles calculated for different values of $\Delta \phi$ with $L=1.5 r_s$, using $\Delta \phi = \Delta \mu$.
The arrows indicate the presence of unphysical charges at the edges of the
numerical grid; (b) same as in (a) but calculated using
our neutrality scheme. Solid line $\Delta \phi/E_F = 0.075$, dashed $\Delta \phi /E_F =0.05$, dotted $\Delta \phi /E_F=0.025$. The vertical lines
indicate the positions of the edges of both jellium surfaces}
\label{Fig-2}
\end{figure}
\begin{figure*}[t]
\includegraphics[scale=0.9]{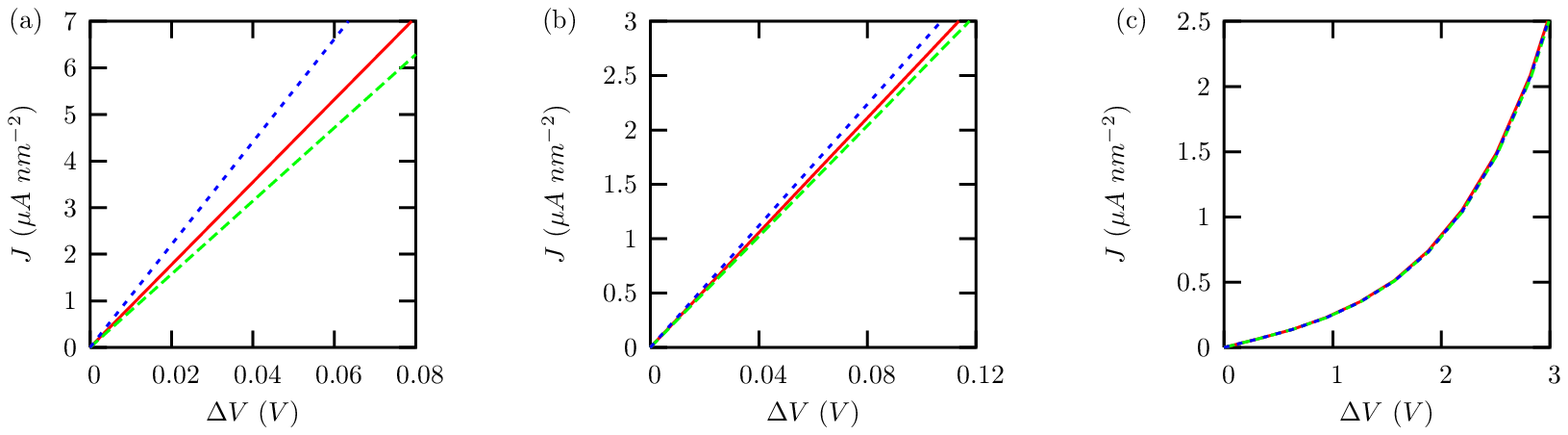}
\caption{(color online) $J-\Delta V$ characteristics for different
electrode-electrode spacings. (a) $L=1 r_s$; (b) $L=1.5 r_s$; (c) $L=3 r_s$. For the solid lines $\Delta V=\Delta \phi=\Delta \mu$, for the dotted lines $\Delta V= \Delta \phi$ and
for the dashed lines $\Delta V=\Delta \mu$}
\label{Fig-3}
\end{figure*}

We next present a set of results for a symmetric metal-vacuum-metal junction with electrodes characterized by $r_s=4$ 
\footnote{Our
conclusions do not depend, qualitatively, on $r_s$}, focusing in
the differences between $\Delta \phi$ and $\Delta \mu$ as a function of the electrode-electrode distance and the effect that the lack
asymptotic neutrality has in the calculated resistivity dipoles and current densities.
Fig. \ref{Fig-1} shows a linear relation between $\Delta \mu$ and $\Delta \phi$ for different lengths of the vacuum gap. All the lines fall
between two limiting ones: $\Delta \phi = 0$ for $L=0$ and $\Delta \phi =\Delta \mu$ for $L \rightarrow \infty$ as expected. For $L=0$
the system is homogeneous along the $z$ direction and hence there is no electrostatic drop. As the distance between electrodes increases the
transmission coefficient $T(E_z)$ decreases, the current decreases and $\rho(z \rightarrow \pm \infty) \rightarrow n_0$,
therefore the electrostatic drop and the difference between electrochemical potentials are approximately equal in that limit. In fact $\Delta
\mu \approx \Delta \phi$ for $L > 2·r_s$. However one should note that there are no molecular conducting channels present in our model.
If these were present and open, then the deviation from $\Delta \mu = \Delta \phi$ should, according to Eq.~(\ref{eq:neutral}), be larger at a fixed value of the
electrode-electrode separation.
Fig.~\ref{Fig-2} shows calculated resistivity dipoles defined as:
\begin{equation}
\delta \rho (z) = \rho(z, \Delta \phi \neq 0)-\rho(z,\Delta \phi = 0),
\end{equation}
for different values of the applied bias. In Fig ~\ref{Fig-2}(a) the dipoles were calculated
within the approximation $\Delta \mu = \Delta \phi$. Enforcing this boundary condition when solving the Poisson equation leads
to the appearance of unphysical charges which are placed at the edges of the numerical grid used in the calculations. These spurious
contributions to the induced density disappear as $\Delta \mu \rightarrow 0$ or $L \rightarrow \infty$, since in these limits $\Delta \mu = \Delta \phi$. 
In Fig ~\ref{Fig-2}(b)  we show the
calculated resistivity dipoles by choosing $k_L$ and $k_R$ so that Eqs.~\ref{eq:asneutral} are satisfied. The induced density
goes smoothly to zero as $z \rightarrow \pm \infty$ even at small values of $L$ and relatively large values of $\Delta \phi$.
Finally in Fig.~\ref{Fig-3} we present calculated $J-\Delta V$ curves (with $\Delta V$ equal to $\Delta \mu$ or $\Delta \phi$ depending on
the pair of curves being compared) for different electrode-electrode spacings. Large difference between the $J-\Delta \phi=\Delta \mu$
curve and any of the other two are present at small electrode-electrode separations. As argued above, as the separation between the electrodes becomes
larger all three curves converge into a single one. Table \ref{tab:G} contains numerical values of the ratios between linear response 
conductances (calculated at a small but finite bias) with and without the approximation $\Delta \mu =\Delta \phi$. The
performance of this approximation is poorer for larger values of the conductance, as expected.
\begin{table}[ht]
\caption{\label{tab:G}Ratios between the calculated linear response conductances per unit area $G_{2P}=J / \Delta \mu$,
$G_{4P}=J/\Delta \phi$ and $G_{\Delta \mu = \Delta \phi}=J/\Delta V$ with $\Delta V=\Delta \mu=\Delta \phi$, together with their
corresponding value of the transmission coefficient evaluated at $E_z=E_F$.}
\begin{tabular}{|r|r|r|r|}
\hline
 $L/r_s$& $G_{2P}/G_{\Delta \mu = \Delta \phi}$ &$G_{4P} / G_{\Delta \mu = \Delta \phi}$&$T(E_z=E_F)$\\
\hline
1.0 & $0.89$& $1.24$&0.675\\
\hline
1.5 & $0.97$&$1.06$&0.347 \\
\hline
3.0 & $1.00$& $1.00$&0.006\\
\hline
\end{tabular}
\end{table}
\section{Relation to the partitioned NEGF approach}
Even though the problem of asymptotic charge neutrality for jellium 
electrodes is clear and directly solvable through the scattering states as 
it is done in this paper, the issue becomes more involved within 
the partitioned NEGF approach. We proceed to show explicitly the terms that
are usually neglected in practical implementations and to provide a simple scheme by
which asymptotic self-consistency may be easily implemented.

Let us consider a partitioned system's Hamiltonian into two semi-infinite
leads ( left (L) and right (R) ) and finite  central region (C).
The Green's functions we consider below are all defined on the Keldysh
contour with the complex time-variable $\tau$~\cite{JauhoBook} and are meant 
to represent matrices with two indexes $m,n$ belonging to some complete set 
of spatially localised basis functions. Each of these can belong to any of 
the above introduced regions L,R or C. 
We employ the notation 
\begin{eqnarray}
	[G_{\alpha}]_{nm} \equiv -i \langle \textrm{T}_{\tau}\{
         \hat{c}_{n \in \alpha}(\tau) \hat{c}^{\dagger}_{m \in \alpha} (\tau')
         \} \rangle \quad \alpha = L,R,C
\end{eqnarray}
for the disconnected systems and 
\begin{eqnarray}
 	[G_{\alpha\beta}]_{nm} \equiv -i \langle \textrm{T}_{\tau}\{ 
	 \hat{c}_{n \in \alpha}(\tau) \hat{c}^{\dagger}_{m \in \beta} (\tau')  
		\} \rangle \quad \alpha,\beta = L,R,C
\end{eqnarray}
for the contacted ones. We also use $1 \equiv \delta_{nm} \delta(\tau-\tau')$.

Before we turn on the contacting between $L$ and $C$ and $C$ and $R$
the uncontacted Green's functions fulfill the equations of motion 
\begin{eqnarray}
	(i\partial_{\tau} - H_L) G_L &=& 1 \\
	(i\partial_{\tau} - H_C) G_C &=& 1 \\
	(i\partial_{\tau} - H_R) G_R &=& 1.
\end{eqnarray}
This can be also written in a block form as
\begin{eqnarray} 
	 \left[ i\partial_{\tau} \bm{1} - \left( \begin{array}{ccc}
		H_L & 0   & 0 \\
		0     & H_C & 0 \\
		0     & 0   & H_{R} 
		\end{array} \right)  \right]
	\left( \begin{array}{ccc}
			G_{L} & 0 & 0 \\
			0 & G_{C} & 0 \\
			0 & 0 & G_{R} 
		\end{array} \right) = \bm 1,
\end{eqnarray}
or more concisely as
\begin{eqnarray} 
	(i \partial_{\tau}  - \bm H ) \bm G^0 = \bm 1. \label{eq-noncontact-G}
\end{eqnarray}

Next we turn on the interaction terms which couple the left and central and 
the right and central parts, written as $V_{L}$ and $V_{R}$ respectively. 
The coupling, however, induces also a change in Hamiltonians via the change 
of the density in the Hartree and the exchange-correlation potentials
which we together write as $\delta H_{\alpha}$ for $\alpha=L,C,R$. 
Using the block Green's functions we have
\begin{equation} 
\label{eq:fullg}
	(i \partial_{\tau}  - \bm H - \delta \bm H - \bm V ) \bm G = \bm 1.
\end{equation}
where
\begin{eqnarray*}
\bm H + \delta \bm H=\left( \begin{array}{ccc}
		H_L^0 + \delta H_L  & V_L & 0 \\
		V^\dagger_L & H_C + \delta H_C & V_{R} \\
		0 & V^\dagger_{R} & H_{R} + \delta H_{R} 
		\end{array} \right),
\end{eqnarray*}
and
\begin{eqnarray*}
\bm G = \left( \begin{array}{ccc}
			G_{LL} & G_{LC} & G_{LR} \\
			G_{CL} & G_{CC} & G_{CR} \\
			G_{RL} & G_{RC} & G_{RR} 
		\end{array} \right).
\end{eqnarray*}
The solution of the Eq.~(\ref{eq:fullg}) can be written using Eq.~(\ref{eq-noncontact-G})
in the form of the Dyson equation as
\begin{equation} 
	\bm G = \bm G^0 + \bm G^0 \left[ \delta \bm H
	 	+ \bm V \right] \cdot \bm G, \label{eq-block-G-dyson}
\end{equation}
where $\cdot$ stands for the integral along the Keldysh contour over an internal 
time variable.

The Green's function of the central region corresponds to the finite system 
and is usually solved numerically in the self-consistent manner. It solves 
the Dyson equation 
\begin{eqnarray}
	G_{CC} &=& G_{C} + G_{C} \delta H_{C} \cdot G_{CC} \nonumber \\ 
	       &+& G_{C} V_{L} \cdot G_{LC} + G_{C} V_{R} \cdot G_{RC} \label{eq-Gcc-dyson}
\end{eqnarray}
which can be found as the '$CC$' component of  
Eq.~(\ref{eq-block-G-dyson}). To have a closed system of equations we need 
to find $G_{LC}$ and $G_{RC}$. These are similarly given as '$LC$' and '$RC$' 
components of Eq.~(\ref{eq-block-G-dyson}) as
\begin{equation} 
	G_{LC} = G_{L} V_{L} \cdot G_{CC} + G_{L} \delta H_{L} \cdot G_{LC},
\end{equation}
and 
\begin{equation} 
	G_{RC} = G_{R} V_{R} \cdot G_{CC} + G_{R} \delta H_{R} \cdot G_{RC}.
\end{equation}
The latter two can be formally inverted (in $m,n$ as well as in $\tau$)  
to give
\begin{eqnarray} 
	G_{LC} &=& \left[1-G_{L} \delta H_{L} \right]^{-1}  \label{eq-Glc-dyson}
		\cdot G_{L} V_{L} \cdot G_{CC} \\
	G_{RC} &=& \left[1-G_{R} \delta H_{R} \right]^{-1}  \label{eq-Grc-dyson}
		\cdot G_{R} V_{R} \cdot G_{CC}.
\end{eqnarray}
Combining Eqs.~(\ref{eq-Gcc-dyson}),(\ref{eq-Glc-dyson}) and (\ref{eq-Grc-dyson})
we finally obtain 
\begin{eqnarray} 
	G_{CC} &=& G_{C} + 
		G_{C} \left( \delta H_{C} +  \Sigma_{CC} \right) \cdot 
		G_{CC},  \\
	\Sigma_{CC} &=& V_{L} \left[1-G_{L} \delta H_{L} \right]^{-1} 
                \cdot G_{L} V_{L} \nonumber \\ 
		&+&
		V_{R} \left[1-G_{R} \delta H_{R} \right]^{-1}  \cdot 
                G_{R} V_{R} \\	
		&=& V_{L} \tilde{G}_{L} V_{L} 
		+ V_{R} \tilde{G}_{R} V_{R},	
\end{eqnarray}
where we have introduced the self-energy $\Sigma_{CC}$ representing the leads 
for the central region and defined new auxiliary Green's functions 
$\tilde{G}_{L/R}$ that fulfill the equations of motion
\begin{equation} 
	\left(i\partial_{\tau} - H_{\alpha} - \delta H_{\alpha}  \right)
	\tilde{G}_{\alpha} = 1, \quad \alpha=L,R. \label{eq-tilde-G}
\end{equation}
The $\tilde{G}_{\alpha}$ need to be such that they correspond to the equilibrium situation 
with $\delta H_{L/R}$ being turn on but with chemical potentials (or Fermi-Dirac
occupation factors $f_{L}$ and $f_{R}$) being kept 
the same as in $G_{L/R}$. For this reason the only significance of
the auxiliary Green's functions is their presence in the expression for 
the self-energy $\Sigma_{CC}$, having no other direct physical 
meaning~\footnote{E.g. the density $\tilde{n} \sim \Im \{ \tilde{G} \}$ 
is different from the physical one.}.
Using the calculated $G_{CC}$ in the Eq.~(\ref{eq-Glc-dyson}) one can now employ 
the usual derivation of the expression for the current in terms of 
$G^{<}_{LC}$~\cite{Meir92}
\begin{equation} 
	I = \frac{2e}{\hbar} \Re\{ \mathrm{Tr}\left[ V G^{<}_{LC} \right] \}.
	\label{eq-I-MW}
\end{equation}

The usual treatments~\cite{Taylor2,Palacios1,Datta1} do not consider 
the change in the leads' Hamiltonians due to the change in density $\delta H_{L/R}$ that arises in the non-equilibrium regime.
This results in a simplification of our equations since the resolvent operators 
$\left[1-G_{R} \delta H_{R} \right]^{-1}$ and 
$\left[1-G_{L} \delta H_{L} \right]^{-1}$ do not have to be calculated in the non-equilibrium regime.
We note that the evaluation of these would require a 
complete calculation of $G_{LL}$ and $G_{RR}$ for the contacted system since these
give the non-equilibrium density that in turn determines the changes 
$\delta H_{L/R}$. This can be most easily achieved for semi-infinite 
jellium electrodes which is equivalent to the calculations presented in the previous section of this 
paper (see also reference \cite{Stefanuci2} for model 1D cases treated 
directly using the non-partitioned NEGF formalism). However completely ignoring these 
changes results in a violation of asymptotic charge neutrality in the leads! 
A practically useful scheme would be to consider this change to be just 
a constant shift, i.e. $\delta H_{L/R} = \delta U_{L/R}$. This is 
exact for jellium electrodes if leads are taken to be sufficiently far 
from the constriction. As it can be seen from Eq.~(\ref{eq-tilde-G}), 
this shift moves the bottoms of the bands in the leads' density of states,
which is the only characteristic of leads that eventually enters into the 
final expression for the current Eq.~(\ref{eq-I-MW}).
The relation between the drop in the chemical 
potentials, the drop in the induced potential and these shifts is simply 
\begin{equation} 
	\Delta \phi + \delta U_{L} - \delta U_{R} = \mu_{R} - \mu_{L},
\end{equation}
which is also clear from Fig.~1c. 
Unfortunately this relation fixes only the differences between the shifts
$\delta U_{L/R}$. We still need to have one more equation to determine them
uniquely. This would come from the computation of non-equilibrium local
charge neutrality in one of the electrodes.
\section{Conclusions}
We have shown that in an exact time-dependent density-functional formulation of the partitioned Keldysh-NEGF approach the 
changes in the Hamiltonian of the leads due to the
contacting needs to be included. It is important for a correct description of the electrostatic potential profile at large currents
or junction with transmission close to one.
Using a simple jellium model of a biased metal-vacuum-metal junction we have examined quantitatively the effects of fixing 
$\Delta \mu =\Delta \phi$.
Significant differences between the non-equilibrium properties calculated using this approximation and 
a more reasonable treatment of the electrostatic
problem based on asymptotic charge neutrality arise in the limit of small electrode-electrode separation. These effects 
would be even more pronounced for resonant molecular junctions where both electrostatics and high conductance acting
simultaneously may significantly influence the $I-V$ characteristics of the system.
\begin{acknowledgments}
The authors gratefully acknowledge useful discussions with
J.J. Palacios. This work was supported by the EU's 6th Framework Programme through 
the NANOQUANTA Network of Excellence (NMP4-CT-2004-500198) and
ERG programme of the European Union QuaTraFo (contract MERG-CT-2004-510615).
\end{acknowledgments}

\begin{thebibliography}{27}
\expandafter\ifx\csname natexlab\endcsname\relax\def\natexlab#1{#1}\fi
\expandafter\ifx\csname bibnamefont\endcsname\relax
  \def\bibnamefont#1{#1}\fi
\expandafter\ifx\csname bibfnamefont\endcsname\relax
  \def\bibfnamefont#1{#1}\fi
\expandafter\ifx\csname citenamefont\endcsname\relax
  \def\citenamefont#1{#1}\fi
\expandafter\ifx\csname url\endcsname\relax
  \def\url#1{\texttt{#1}}\fi
\expandafter\ifx\csname urlprefix\endcsname\relax\def\urlprefix{URL }\fi
\providecommand{\bibinfo}[2]{#2}
\providecommand{\eprint}[2][]{\url{#2}}

\bibitem[{\citenamefont{McCann and Brown}(1988)}]{McCannBrown}
\bibinfo{author}{\bibfnamefont{A.}~\bibnamefont{McCann}} \bibnamefont{and}
  \bibinfo{author}{\bibfnamefont{J.~S.} \bibnamefont{Brown}},
  \bibinfo{journal}{Surf. Sci.} \textbf{\bibinfo{volume}{194}},
  \bibinfo{pages}{44} (\bibinfo{year}{1988}).

\bibitem[{\citenamefont{Lang}(1995)}]{Lang1}
\bibinfo{author}{\bibfnamefont{N.~D.} \bibnamefont{Lang}},
  \bibinfo{journal}{Phys.Rev. B} \textbf{\bibinfo{volume}{52}},
  \bibinfo{pages}{5335} (\bibinfo{year}{1995}).

\bibitem[{\citenamefont{Damle et~al.}(2001)\citenamefont{Damle, Ghosh, and
  Datta}}]{Datta1}
\bibinfo{author}{\bibfnamefont{P.~S.} \bibnamefont{Damle}},
  \bibinfo{author}{\bibfnamefont{A.~W.} \bibnamefont{Ghosh}}, \bibnamefont{and}
  \bibinfo{author}{\bibfnamefont{S.}~\bibnamefont{Datta}},
  \bibinfo{journal}{Phys. Rev. B} \textbf{\bibinfo{volume}{64}},
  \bibinfo{pages}{201403(R)} (\bibinfo{year}{2001}).

\bibitem[{\citenamefont{Taylor et~al.}(2001)\citenamefont{Taylor, Guo, and
  Wang}}]{Taylor1}
\bibinfo{author}{\bibfnamefont{J.}~\bibnamefont{Taylor}},
  \bibinfo{author}{\bibfnamefont{H.}~\bibnamefont{Guo}}, \bibnamefont{and}
  \bibinfo{author}{\bibfnamefont{J.}~\bibnamefont{Wang}},
  \bibinfo{journal}{Phys.Rev. B} \textbf{\bibinfo{volume}{63}},
  \bibinfo{pages}{245407} (\bibinfo{year}{2001}).

\bibitem[{\citenamefont{Brandbyge et~al.}(2002)}]{Taylor2}
\bibinfo{author}{\bibfnamefont{M.}~\bibnamefont{Brandbyge}}
  \bibnamefont{et~al.}, \bibinfo{journal}{Phys. Rev. B}
  \textbf{\bibinfo{volume}{65}}, \bibinfo{pages}{165401}
  (\bibinfo{year}{2002}).

\bibitem[{\citenamefont{Palacios et~al.}(2002)}]{Palacios1}
\bibinfo{author}{\bibfnamefont{J.J.}~\bibnamefont{Palacios}}
  \bibnamefont{et~al.}, \bibinfo{journal}{Phys. Rev. B}
  \textbf{\bibinfo{volume}{66}}, \bibinfo{pages}{035322}
  (\bibinfo{year}{2002}).

\bibitem[{\citenamefont{Louis et~al.}(2003)}]{Palacios2}
\bibinfo{author}{\bibfnamefont{E.}~\bibnamefont{Louis}} \bibnamefont{et~al.},
  \bibinfo{journal}{Phys. Rev. B} \textbf{\bibinfo{volume}{67}},
  \bibinfo{pages}{155321} (\bibinfo{year}{2003}).

\bibitem[{\citenamefont{Kohn and Sham}(1965)}]{KS}
\bibinfo{author}{\bibfnamefont{W.}~\bibnamefont{Kohn}} \bibnamefont{and}
  \bibinfo{author}{\bibfnamefont{L.~J.} \bibnamefont{Sham}},
  \bibinfo{journal}{Phys. Rev.} \textbf{\bibinfo{volume}{140}},
  \bibinfo{pages}{A1133} (\bibinfo{year}{1965}).

\bibitem[{\citenamefont{Hohenberg and Kohn}(1964)}]{HK}
\bibinfo{author}{\bibfnamefont{P.}~\bibnamefont{Hohenberg}} \bibnamefont{and}
  \bibinfo{author}{\bibfnamefont{W.}~\bibnamefont{Kohn}},
  \bibinfo{journal}{Phys. Rev.} \textbf{\bibinfo{volume}{136}},
  \bibinfo{pages}{B864} (\bibinfo{year}{1964}).

\bibitem[{\citenamefont{Keldysh}(1964)}]{Keldysh}
\bibinfo{author}{\bibfnamefont{L.~V.} \bibnamefont{Keldysh}},
  \bibinfo{journal}{Sov. Phys. JETP} \textbf{\bibinfo{volume}{20}},
  \bibinfo{pages}{1018} (\bibinfo{year}{1964}).

\bibitem[{\citenamefont{Caroli et~al.}(1972)}]{Caroli}
\bibinfo{author}{\bibfnamefont{C.}~\bibnamefont{Caroli}} \bibnamefont{et~al.},
  \bibinfo{journal}{J. Phys. C} \textbf{\bibinfo{volume}{5}},
  \bibinfo{pages}{21} (\bibinfo{year}{1972}).

\bibitem[{\citenamefont{Hirose and Tsukada}(1995)}]{Hirose}
\bibinfo{author}{\bibfnamefont{K.}~\bibnamefont{Hirose}} \bibnamefont{and}
  \bibinfo{author}{\bibfnamefont{M.}~\bibnamefont{Tsukada}},
  \bibinfo{journal}{Phys. Rev. B} \textbf{\bibinfo{volume}{51}},
  \bibinfo{pages}{5278} (\bibinfo{year}{1995}).

\bibitem[{\citenamefont{Cini}(1980)}]{Cini80}
\bibinfo{author}{\bibfnamefont{M.}~\bibnamefont{Cini}}, \bibinfo{journal}{Phys.
  Rev. B} \textbf{\bibinfo{volume}{22}}, \bibinfo{pages}{5887}
  (\bibinfo{year}{1980}).

\bibitem[{\citenamefont{Stefanucci and
  Almbladh}(2004{\natexlab{a}})}]{Stefanuci1}
\bibinfo{author}{\bibfnamefont{G.}~\bibnamefont{Stefanucci}} \bibnamefont{and}
  \bibinfo{author}{\bibfnamefont{C.-O.} \bibnamefont{Almbladh}},
  \bibinfo{journal}{Europhys. Lett.} \textbf{\bibinfo{volume}{67}},
  \bibinfo{pages}{14} (\bibinfo{year}{2004}{\natexlab{a}}).

\bibitem[{\citenamefont{Stefanucci and
  Almbladh}(2004{\natexlab{b}})}]{Stefanuci2}
\bibinfo{author}{\bibfnamefont{G.}~\bibnamefont{Stefanucci}} \bibnamefont{and}
  \bibinfo{author}{\bibfnamefont{C.-O.} \bibnamefont{Almbladh}},
  \bibinfo{journal}{Phys. Rev. B} \textbf{\bibinfo{volume}{69}},
  \bibinfo{pages}{195318} (\bibinfo{year}{2004}{\natexlab{b}}).

\bibitem[{\citenamefont{Landauer}(1957)}]{Landauer57}
\bibinfo{author}{\bibfnamefont{R.}~\bibnamefont{Landauer}},
  \bibinfo{journal}{IBM J. Res. Dev} \textbf{\bibinfo{volume}{1}},
  \bibinfo{pages}{223} (\bibinfo{year}{1957}).

\bibitem[{\citenamefont{B{\"u}ttiker et~al.}(1985)\citenamefont{B{\"u}ttiker,
  Imry, Landauer, and Pinhas}}]{Buttiker85}
\bibinfo{author}{\bibfnamefont{M.}~\bibnamefont{B{\"u}ttiker}},
  \bibinfo{author}{\bibfnamefont{Y.}~\bibnamefont{Imry}},
  \bibinfo{author}{\bibfnamefont{R.}~\bibnamefont{Landauer}}, \bibnamefont{and}
  \bibinfo{author}{\bibfnamefont{S.}~\bibnamefont{Pinhas}},
  \bibinfo{journal}{Phys. Rev. B} \textbf{\bibinfo{volume}{31}},
  \bibinfo{pages}{6207} (\bibinfo{year}{1985}).

\bibitem[{\citenamefont{P{\"o}tz}(1989)}]{potz}
\bibinfo{author}{\bibfnamefont{W.}~\bibnamefont{P{\"o}tz}},
  \bibinfo{journal}{J. Appl. Phys.} \textbf{\bibinfo{volume}{66}},
  \bibinfo{pages}{2458} (\bibinfo{year}{1989}).

\bibitem[{\citenamefont{Bokes and Godby}(2003)}]{Bokes}
\bibinfo{author}{\bibfnamefont{P.}~\bibnamefont{Bokes}} \bibnamefont{and}
  \bibinfo{author}{\bibfnamefont{R. W.}~\bibnamefont{Godby}},
  \bibinfo{journal}{Phys. Rev. B} \textbf{\bibinfo{volume}{68}},
  \bibinfo{pages}{125414} (\bibinfo{year}{2003}).

\bibitem[{\citenamefont{Liang et~al.}(2004)\citenamefont{Liang, Ghosh, Paulsson,
  and Datta}}]{Datta2}
\bibinfo{author}{\bibfnamefont{G. C.}~\bibnamefont{Liang}},
  \bibinfo{author}{\bibfnamefont{A. W.}~\bibnamefont{Ghosh}},
  \bibinfo{author}{\bibfnamefont{M.}~\bibnamefont{Paulsson}}, \bibnamefont{and}
  \bibinfo{author}{\bibfnamefont{S.}~\bibnamefont{Datta}},
  \bibinfo{journal}{Phys. Rev. B} \textbf{\bibinfo{volume}{69}},
  \bibinfo{pages}{115302} (\bibinfo{year}{2004}).

\bibitem[{\citenamefont{Ferry and Goodnick}(1997)}]{ferry}
\bibinfo{author}{\bibfnamefont{D.~K.} \bibnamefont{Ferry}} \bibnamefont{and}
  \bibinfo{author}{\bibfnamefont{S.~M.} \bibnamefont{Goodnick}},
  \emph{\bibinfo{title}{Transport in Nanostructures}}
  (\bibinfo{publisher}{Cambridge Univ. Press, Cambridge, UK},
  \bibinfo{year}{1997}).

\bibitem[{\citenamefont{Ventra and Lang}(2001)}]{DiventraLang}
\bibinfo{author}{\bibfnamefont{M.} \bibnamefont{DiVentra}} \bibnamefont{and}
  \bibinfo{author}{\bibfnamefont{N.~D.} \bibnamefont{Lang}},
  \bibinfo{journal}{Phys. Rev. B} \textbf{\bibinfo{volume}{65}},
  \bibinfo{pages}{045402} (\bibinfo{year}{2001}).

\bibitem[{\citenamefont{Nieminen}(1977)}]{Nieminen}
\bibinfo{author}{\bibfnamefont{R.~M.} \bibnamefont{Nieminen}},
  \bibinfo{journal}{J. Phys. F: Metal Phys.} \textbf{\bibinfo{volume}{7}},
  \bibinfo{pages}{375} (\bibinfo{year}{1977}).

\bibitem[{\citenamefont{Orosz and Balazs}(1986)}]{OroszBalazs}
\bibinfo{author}{\bibfnamefont{L.}~\bibnamefont{Orosz}} \bibnamefont{and}
  \bibinfo{author}{\bibfnamefont{E.}~\bibnamefont{Balazs}},
  \bibinfo{journal}{Surf. Sci.} \textbf{\bibinfo{volume}{177}},
  \bibinfo{pages}{444} (\bibinfo{year}{1986}).

\bibitem[{\citenamefont{Perdew and Zunger}(1981)}]{PerdewZunger}
\bibinfo{author}{\bibfnamefont{J.~P.} \bibnamefont{Perdew}} \bibnamefont{and}
  \bibinfo{author}{\bibfnamefont{A.}~\bibnamefont{Zunger}},
  \bibinfo{journal}{Phys. Rev. B} \textbf{\bibinfo{volume}{23}},
  \bibinfo{pages}{5048} (\bibinfo{year}{1981}).

\bibitem[{\citenamefont{Haug and Jauho}(1996)}]{JauhoBook}
\bibinfo{author}{\bibfnamefont{H.}~\bibnamefont{Haug}} \bibnamefont{and}
  \bibinfo{author}{\bibfnamefont{A.-P.} \bibnamefont{Jauho}},
  \emph{\bibinfo{title}{Quantum Kinetics in Transport and Optics of
  Semiconductors}} (\bibinfo{publisher}{Springer Verlag},
  \bibinfo{year}{1996}).

\bibitem[{\citenamefont{Meir and Wingreen}(1992)}]{Meir92}
\bibinfo{author}{\bibfnamefont{Y.}~\bibnamefont{Meir}} \bibnamefont{and}
  \bibinfo{author}{\bibfnamefont{N. S.}~\bibnamefont{Wingreen}},
  \bibinfo{journal}{Phys. Rev. Lett.} \textbf{\bibinfo{volume}{68}},
  \bibinfo{pages}{2512} (\bibinfo{year}{1992}).

\end{thebibliography}

\end{document}